\documentclass[11pt]{article}
\usepackage{epsf,amssymb,cite}
\newcommand\ttl[1]{`#1', }
\newcommand\hepth[1]{, {\tt hep-th/#1}}
\newcommand\heplat[1]{, {\tt hep-lat/#1}}
\newcommand\condmat[1]{, {\tt cond-mat/#1}}
\newcommand\Zt{{\mathbb Z}_2}
\newcommand\Zth{{\mathbb Z}_3}
\newcommand\Zf{{\mathbb Z}_5}
\newcommand\Zs{{\mathbb Z}_6}
\newcommand\ZN{{\mathbb Z}_N}
\newcommand\eq{\begin{equation}}
\newcommand\en{\end{equation}}
\newcommand\bea{\begin{eqnarray}}
\newcommand\eea{\end{eqnarray}}
\newcommand\nn{\nonumber}
\newcommand\half{{\textstyle\frac{1}{2}}}
\newcommand\Epsilon{{\cal E}}
\begin{document}
\begin{titlepage}
\vskip 0.5cm
\begin{flushright}
DTP-98/75 \\
DFTT-62/98 \\
T98/109 \\
MS-TPI-98-21 \\
{\tt hep-th/9810202} \\
October 1998
\end{flushright}
\vskip 1.2cm
\begin{center}
{\Large{\bf On the phase diagram of the discrete $\Zs$ spin models}}
\end{center}
\vskip 0.8cm
\centerline{Patrick Dorey%
\footnote{e-mail: {\tt p.e.dorey@durham.ac.uk}},
Paolo Provero%
\footnote{e-mail: {\tt provero@to.infn.it}},
Roberto Tateo%
\footnote{e-mail: {\tt tateo@wasa.saclay.cea.fr}}
and Stefano Vinti%
\footnote{e-mail: {\tt  vinti@uni-muenster.de}}}
\vskip 0.9cm
\centerline{${}^1$\sl\small Department of Mathematical Sciences,
University of Durham, Durham DH1 3LE, UK\,}
\vskip 0.2cm
\centerline{${}^2$\sl\small Dipartimento di Fisica Teorica dell'Universit\`a
di Torino, via P.Giuria 1, I-10125  Torino, Italy\,}
\vskip 0.2cm
\centerline{${}^3$\sl\small Service de Physique Th\'eorique, CEA-Saclay,
F-91191 Gif-sur-Yvette Cedex, France\,}
\vskip 0.2cm
\centerline{${}^4$\sl\small Institut f\"ur Theoretische Physik I,
Universit\"at M\"unster, D-48149 M\"unster, Germany\,}
\vskip 1.2cm
\begin{abstract}
We point out some problems with the previously-proposed phase diagram
of the $\Zs$ spin models. Consideration of the diagram near to the
decoupling surface using both exact and approximate arguments suggests
a modification which remedies these deficiencies. With the aid of a new
parametrisation of the phase space, we study the models numerically,
with results which support our conjectures.
\end{abstract}
\end{titlepage}
\setcounter{footnote}{0}
\def\thefootnote{\fnsymbol{footnote}}
The discrete $\Zs$ spin models describe the behaviour of
a collection of spins $S_i=\exp(i\theta_i)$, where the $\theta_i$
are
integer multiples of $\pi/3$. These spins
live on a square two-dimensional lattice and
interact according to a reduced Hamiltonian 
of the form
\eq
H\equiv{\textstyle\frac{1}{k_BT}}{\cal H}=
\sum_{<ij>}V(\theta_i-\theta_j)\,,
\label{ham}
\en
the sum running over nearest-neighbour pairs of sites $<ij>$.
Imposing $V(\theta)=V(-\theta)$, a particular system is
characterised by the three numbers $V_r=V(\pi r/3)-V(0)$, $r=1,2,3$.
A duality transformation maps the space of such models to 
itself\cite{WWa,DRa,Ca,AKa}: with the couplings parametrised by
variables $x_r=\exp(-V_r)$, a sum over the $\theta_i$ is
equivalent to one over dual variables $\tilde\theta_{\tilde\imath}$, with
the triplet of couplings $\{x_r\}$ replaced by the dual set $\{\tilde
x_r\}$:
\bea
\tilde x_1&=&(1\,+\,x_1\,-\,x_2\,-x_3)/\Delta \nn\\
\tilde x_2&=&(1\,-\,x_1\,-\,x_2\,+x_3)/\Delta \nn\\
\tilde x_3&=&(1-2x_1+2x_2-x_3)/\Delta
\label{duality}
\eea
where $\Delta=1+2x_1+2x_2+x_3$.
The transformation leaves invariant points on the line 
\eq(x_1,x_2,x_3)=(t,\beta{-}\alpha t,\alpha{-}2\beta t)
\label{sdline}
\en
where $\alpha=3{-}\sqrt{6}$ and $\beta=\sqrt6{-}2$. This
self-dual line intersects the line of 6-state Potts
models at the point
$x_1=x_2=x_3=(\sqrt{6}{-}1)/5$, denoted $P$ below. 
There is also a distinguished surface
on which the models decouple into independent
three-state Potts and Ising 
models\cite{DRa}. This is revealed on writing each spin $S_i$ as a
product of a $\Zt$ and a $\Zth$ valued variable:
\eq
S_i=e^{i\theta_i}=\Sigma_i\sigma_i\,;\quad~ \Sigma_i=1,e^{\pm2\pi i/3},
\quad \sigma=\pm 1
\en
The pairwise interaction energy can then be written as
\bea
V(\theta)&=&J_1[1{-}\cos(\theta)]+J_2[1{-}\cos(2\theta)]
+J_3[1{-}\cos(3\theta)]\nn\\[3pt]
&=&
J_1[1{-}\half(S{+}S^{-1})]
+J_2[1{-}\half(\Sigma{+}\Sigma^{-1})]
+ J_3[1{-}\half(\sigma{+}\sigma^{-1})]\nn\\[3pt]
&=&
3J_1[1{-}\delta_{\sigma_i\sigma_j}\delta_{\Sigma_i\Sigma_j}]+
{\textstyle\frac{3}{2}}(J_2{-}J_1)[1{-}\delta_{\Sigma_i\Sigma_j}]
+ (2J_3{-}J_1)[1{-}\delta_{\sigma_i\sigma_j}]
\label{decoup}
\eea
where $\theta=\theta_i{-}\theta_j$,
$S=S_i/S_j$, $\Sigma=\Sigma_i/\Sigma_j$,
$\sigma=\sigma_i/\sigma_j$, and
\eq
\textstyle
J_1=\frac{1}{3}\ln\frac{x_1}{x_2x_3}\quad
J_2=\frac{1}{3}\ln\frac{x_3}{x_1x_2}\quad
J_3=\frac{1}{6}\ln\frac{x_2^2}{x_1^2x_3}~.
\en
The spins $\Sigma_i$ and $\sigma_i$ decouple on the surface
$J_1=0$, $x_1=x_2x_3$,
on which $\frac{3}{2}J_2$ is the 3-state Potts coupling, and $2J_3$
the Ising coupling.

The more general
phase structure of these models has been studied by various authors
over the years (see for example \cite{DRa,Ca,AKa,GRa}). 
The question is an interesting exercise in its own
right, and is also of wider relevance -- to, for example, the
effect of hexagonal symmetry-breaking on the isotropic planar model
\cite{JKKNa,DRa}, 
the behaviour of the cubic model \cite{KLUa,DRa}, and the spectra of
Heisenberg antiferromagnetic spin chains \cite{CPRa}.
An initial phase diagram was suggested by Domany and Riedel in
\cite{DRa}, with the three-dimensional thermodynamic phase space
partitioned into four domains, one disordered and the others
exhibiting $\Zt$, $\Zth$ and $\Zs$ ordering. The existence of an
additional massless phase was then demonstrated, first\cite{EPSa}
along the Villain\cite{Va} line, and then\cite{Ca} throughout a whole
three-dimensional region. This phase was incorporated into the
Domany-Riedel diagram by Alcaraz and Koberle in\cite{AKa}, its
end point on the self-dual line later being identified with a particular
integrable point $C$ on the phase diagram\cite{FZa}
(further support for
such an identification can be found in\cite{DMSa,DTTa}).

However there are reasons to believe that the story is not yet
complete. We give two examples.

\noindent (i) In the diagrams of\cite{DRa} and\cite{AKa}, the 
point $P$ touches
surfaces of transition from the disordered phase into regions of $\Zt$
and $\Zth$ order. Such transitions are expected to be of second order,
and so the correlation length on these surfaces should be infinite.
This contradicts the known behaviour of the 6-state Potts model at
$P$,
where the correlation length remains finite\cite{Ba},
albeit large\cite{BWa}.

\noindent (ii) On the decoupling surface $J_1=0$, there is a line of
Ising transitions, $J_3=\half\ln[1{+}\sqrt{2}]$, and a line of
3-state Potts transitions, 
$J_2=\frac{2}{3}\ln[1{+}\sqrt{3}]$. These lines cross at 
a renormalisation group fixed point $D$, the product
of a critical Ising model and a critical three-state Potts model.
Three $\Zs$-invariant operators at this point are the Ising and
three-state Potts energy densities $\epsilon$ and $\Epsilon$, and
their product $\epsilon\Epsilon$. Since their scaling dimensions are
$1$, $4/5$ and $9/5$ respectively, all are relevant and the fixed
point is triply unstable. This corrects the approximate (Migdal)
renormalisation group result used in\cite{DRa}, which gave $D$ as being
doubly-unstable.

We can take this second point a little further, using continuum field
theory arguments which are valid in the scaling region around the
point $D$. The fixed point itself is described by a product of $c=1/2$
and $c=4/5$ conformal field theories, and nearby points by
perturbations of this product $c=13/10$
conformal field theory by combinations
of the continuum operators $\epsilon$, $\Epsilon$ and
$\epsilon\Epsilon$. The first two are anti-self-dual under 
(\ref{duality}), while the third is self-dual and moves the model away
from $D$ along the line (\ref{sdline}). 
Minimal models coupled by local operators have received a
fair amount of attention (see, for example, \cite{Caa}) but this
particular instance does not seem to have been studied in any detail. 
However,
Zamolodchikov's counting argument~\cite{Za} can be used to
show that the $\epsilon\Epsilon$ perturbation preserves at the very least
conserved charges of spins $\pm 3$ and $\pm 5$. The 
perturbed (continuum) theory should
therefore be integrable~\cite{Pa} and
we can hope to obtain 
information about the scaling region of the self-dual line near to
$D$ via the thermodynamic Bethe ansatz
(TBA) technique\cite{Zb}. This method expresses the finite-volume `effective
central charge' $c(r)$, $r=mR$, of a model with bulk 
length scale
$1/m$ confined to a circle of circumference $R$ in terms of the solutions
$\varepsilon_a(\theta)$ to a set of coupled integral equations.
A candidate system for this case has already been identified:
in\cite{RTVa} it was observed that the following TBA
system for the functions $\varepsilon_1\dots\varepsilon_7$
\bea
\varepsilon_a(\theta)&=&\delta_{a1}r\cosh\theta- \frac{1}{2 \pi}
\sum_{b=1}^7 l^{[E_7]}_{ab} 
 \phi* \ln(1{+}e^{-\varepsilon_b})(\theta)~,\nn\\
c(r)&=&\frac{3}{\pi^2}\int_{-\infty}^{\infty}\! d\theta\,
r\cosh\theta\ln(1{+}e^{-\varepsilon_1(\theta)})
\label{TBA}
\eea
predicts an ultraviolet central charge of $13/10$, and a small-$r$ behaviour
of $c(r)$ compatible with an expansion in even powers of a coupling
$\lambda$ to an operator of dimension $9/5$.
(Here $*$ denotes the convolution,
$f{*}g(\theta)=
\int^{\infty}_{-\infty}\!d\theta' f(\theta{-}\theta')g(\theta')$, 
$\phi(\theta)=1/\cosh\theta$, and $l^{[E_7]}_{ab}$ is the incidence matrix of
the $E_7$ Dynkin diagram, with $1$ labeling the node at the end of the
middle-length arm.)
This is the behaviour expected of
the perturbation of the product Ising
and three-state Potts conformal field theory 
by its $\epsilon\Epsilon$ operator. 
(On dimensional grounds, $m$ will be related to $\lambda$ as
$m\propto\lambda^5$, but we do not attempt to find the constant of
proportionality here.)
Assuming that the TBA is correct, it can be used to
extract a non-trivial prediction about the vacuum structure of the
perturbed model, using arguments described in section 4.3
of\cite{DTTa}. The system (\ref{TBA})
implies the following asymptotic for the ground-state energy $E(m,R)$:
\eq
E(m,R)=-{ \pi \over 6 R}c(mR) \sim -\frac{\sqrt{6}\,m}{\pi} K_1(mR)
\label{TBAasymptotic}
\en
where $K_1$ is the modified Bessel function of order one and 
the prefactor $\sqrt{6}$ can be interpreted as the 
Perron-Frobenius eigenvalue of
the incidence matrix for single kinks interpolating degenerate vacua
of the perturbed theory\cite{DTTa}. These vacua must support a
representation of the global $\Zs$ symmetry, and we find 
that the simplest incidence matrices
compatible with both this fact and the eigenvalue implied by
(\ref{TBAasymptotic}) are the following:
\eq
I^a=
\left(\begin{tabular}{ccccccc}
0&1&1&1&1&1&1\\
1&0&0&0&0&0&0\\
1&0&0&0&0&0&0\\
1&0&0&0&0&0&0\\
1&0&0&0&0&0&0\\
1&0&0&0&0&0&0\\
1&0&0&0&0&0&0
\end{tabular}\right)
\quad
I^b=
\left(\begin{tabular}{ccccc}
0&0&1&1&1\\
0&0&1&1&1\\
1&1&0&0&0\\
1&1&0&0&0\\
1&1&0&0&0
\end{tabular}\right)
\en
The first is consistent with the coexistence of disordered
and $\Zs$ ordered phases, the second with coexistence of $\Zt$ and
$\Zth$ ordered phases. This leads us to the first part of our proposal
for the resolution of points (i) and (ii) above: we suggest that the 
entire segment of the self-dual line from $D$ to $C$, which includes the
6-state Potts transition point $P$, lies on a surface of first-order
transitions separating disordered and fully-ordered regions, and
in the scaling region near to $D$ is described by an integrable
scattering theory with kink incidence matrix $I^a$. (In the massive
scaling region near to $C$, the vacuum structure comes as
the $N{=}6$ case of the results of\cite{DTTa}, and is
also given by $I^a$.)
It is natural to suppose that the matrix $I^b$ describes
the vacuum structure on the opposite side of the point $D$, and
support for this idea comes from the following argument.

Consider the behaviour of the model near to the
decoupling surface $J_1=0$, taking either $J_2\gg 1$ or
$J_3\gg 1$. In the first case, the effect will be to freeze
out the three-state Potts spins $\Sigma_i$,
and all terms $\delta_{\Sigma_i\Sigma_j}$ can be approximated by
$1$ in the final line of (\ref{decoup}). The result is an
effective interaction for the remaining unfrozen spins, 
equal to
$2(J_3{+}J_1)[1{-}\delta_{\sigma_i\sigma_j}]$. Thus the only
effect of a non-zero $J_1$ in this `$\Sigma$-frozen' region
is to replace $J_3$ with $J_3{+}J_1$.
The line of Ising transitions on the decoupling surface was at
$J_3=J_3^c=\half\ln[1{+}\sqrt{2}]\,$; we now conclude that
a small
non-zero $J_1$ in the region $J_2\gg 1$
simply shifts this line to $J_3=J_3^c{-}J_1$.
A similar argument shows that the line of three-state Potts transitions,
situated at $J_2=J_2^c=\frac{2}{3}\ln[1{+}\sqrt{3}]$ on the decoupling
surface, is shifted to $J_2=J_2^c{-}J_1$ in the region $J_3\gg 1$
where the $\sigma$ spins are frozen. Finally, duality can
be used to see that in the opposite regimes $J_2\ll 1$ and $J_3\ll
1$, the critical lines move in the opposite senses when $J_1$ shifts
away from zero. This line of argument has nothing to say directly
about the central region where the Ising and three-state
Potts spins are both near their critical points, but the simplest
hypothesis, also consistent with the continuum results of the
last paragraph, is to continue the lines in from the asymptotic
regions as in Fig.~1.

\begin{figure}[h]
\vskip 0.3cm
\centerline{\epsfxsize=.74\linewidth\epsfbox{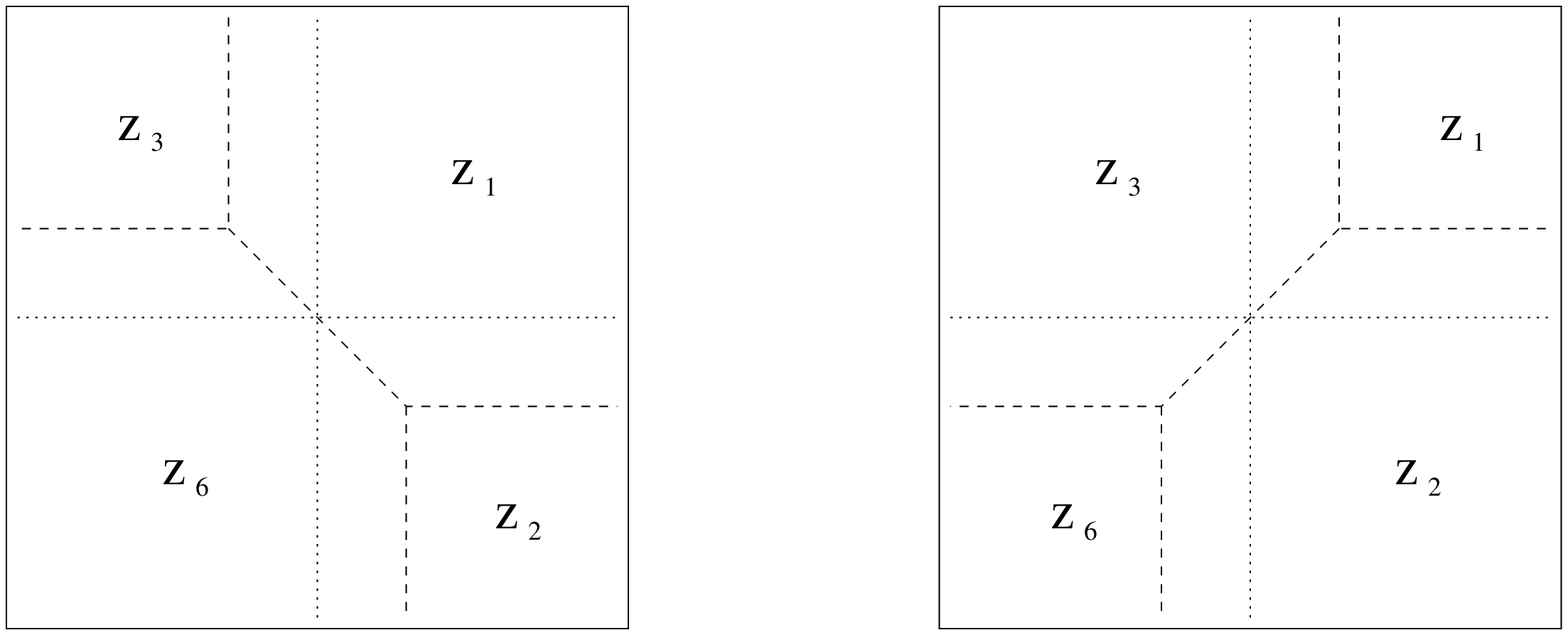}}
\vskip 0.15cm
\centerline{\parbox[t]{0.84\linewidth}{
 \small Figure 1: Schematic pictures of slices through the phase
diagram just below, and just above, the decoupling surface.}}
\label{fig1}
\end{figure}

Much of our numerical work has been devoted to the confirmation of
this picture. Before we describe this, we pause to
introduce a third set of coordinates on the thermodynamic phase space.
First, notice that the parameter $K$ defined by
\eq
K(x_1,x_2,x_3)=(x_1{-}x_2x_3)/\Delta
\en
is mapped into itself by duality: $\tilde{K}=K\,$.
Therefore surfaces of constant $K$ in phase space are mapped into themselves
by duality; $K=0$ is the decoupling surface.  Points with $K>0$
lie `below' this surface, on the same side as the 6-state Potts
point $P$. 
It is easy to check that
the two parameters
\eq
y_2=\ln\frac{1{+}2x_2}{\sqrt{3{-}6K}}\qquad
y_3=\ln\frac{1{+}x_3}{\sqrt{2{-}4K}} 
\en
have the simple duality transformation rule
\eq
\tilde{y_2}=-y_2\qquad\tilde{y_3}=-y_3\quad
\en
It is convenient to use $(K,y_2,y_3)$ as coordinates in phase space: on any 
fixed-$K$ surface the point $(K,0,0)$ is the self-dual point and duality
is just the reflection in this point. The origin $(0,0,0)$ 
is the fixed point $D$ on the decoupling surface, and the 6-state
Potts transition point $P$ is at $(K_P,0,0)$ with
$K_P=(\sqrt{6}{-}1)^2/25\approx 0.084$.
On the decoupling surface, $y_2$ and $y_3$ are functions of the Ising and
Potts couplings respectively.
\par
A cluster algorithm appears to be the ideal choice to simulate 
the relevant regions of the phase diagram,
since it is reasonable to expect 
large correlation lengths even when the transitions are only first-order. 
Since
cluster algorithms are especially simple to implement for $\Zt$ and $\Zth$
models, we exploited the possibility of writing the Hamiltonian as in 
(\ref{decoup}): the algorithm we used performs alternate cluster updates of 
the $\sigma_i$ and $\Sigma_i$ variables, where those which are not being 
updated provide effective, site-dependent couplings for the others.
For example, suppose we are updating the $\Zt$ variables: the effective 
coupling to be used on the link $<ij>$ is then
$J_3+J_1\left(\Sigma{+}\Sigma^{-1}\right)/2$
with $\Sigma=\Sigma_i/\Sigma_j$,
while when updating the $\Zth$ variables the effective coupling is
$J_2+J_1\sigma$
with $\sigma=\sigma_i/\sigma_j$.
Similar algorithms were introduced in \cite{ashkinteller} for the Ashkin 
Teller model, defined as two coupled Ising models.   
\par
The fact that the effective couplings are site-dependent does not pose
a problem for the cluster updates. 
However the effective $\Zt$ and $\Zth$
couplings can become negative, {\em i.e.}\ antiferromagnetic. 
This will happen for the effective $\Zt$ coupling whenever
$J_3+J_1< 0$ or $J_3-J_1/2< 0$, and for the $\Zth$ coupling
whenever $J_2+J_1< 0$ or $J_2-J_1< 0$.
Negative couplings 
on some links can in turn lead to frustrations, which in principle can 
make the cluster algorithm highly ineffective.
However it is easy to convince oneself that 
the $\Zth$ model is never actually
frustrated: for every configuration of the $\Zt$ spins there exists a
configuration of the $\Zth$ ones such that all the links are satisfied.
Therefore the possibility of frustrations exists only for $J_1<-J_3$ or
$J_1>2 J_3$. These relations are never satisfied in the regions we
considered.
\par  
We explored several fixed-$K$ surfaces 
and mapped the various phases and their boundaries using the Binder 
cumulants method \cite{binder}.
In our case, we define two cumulants, one for each order 
parameter:
\eq
Q_\sigma=\frac{3}{2}-\frac{1}{2} \,\frac{\langle m_\sigma^4\rangle}
{\langle m_\sigma^2\rangle^2}\qquad
Q_\Sigma=2- \,\frac{\langle m_\Sigma^4\rangle}
{\langle m_\Sigma^2\rangle^2}
\en
where $m_\sigma$ and $m_\Sigma$ are the $\Zt$ and $\Zth$ magnetizations
per site. 
For each surface we measured the $\Zt$ and $\Zth$ Binder cumulants on a grid
of points around the self-dual point $(K,0,0)$ for two different lattice 
sizes $L_1{\times}L_1$ and $L_2{\times}L_2$. 
For every fixed value of $y_2$ we estimated the value
of $y_3$ where each Binder cumulant remains constant when the lattice size is
increased, {\em i.e.}\ the transition point. 
This allows us to determine a set of points belonging to
the transition lines for the $\Zt$ and $\Zth$ variables. These are the points
plotted in Figs.~2,3,4 for the $K=0.06,\ 0.04,\  -0.08$ 
surfaces respectively. The consistency of the procedure was checked by 
repeating it with the roles of $y_2$ and $y_3$ inverted.   
Finite-size effects are signaled by violations of the duality symmetry, that
is, in our coordinates, symmetry under reflection in the origin. Small
violations are visible in our figures; we checked that they become smaller when
the lattice sizes are increased. 
These effects turned out to be
much larger in the $K<0$ region.
In fact we observed that to 
control them satisfactorily
in the $K>0$ region lattice 
sizes as small as $L_1=15$, $L_2=30$ sufficed, while for 
$K<0$ we had to use 
$L_1=45$, $L_2=60$.  
\par
The numerical results confirm the qualitative picture shown
in Fig.~1, and our claim that the decoupling point
$D$ on the self-dual line marks a change from the coexistence of
disorder and $\Zs$ order to the coexistence of $\Zt$ and $\Zth$ order.
The reduced length of the first-order segment at $K=0.04$ is
consistent with scaling predictions, though we are not close
enough to the point $D$ to see a complete collapse of data.
In conclusion, we have proposed a significant modification to the
previously-accepted phase diagram of the $\Zs$ spin models, and
this has been supported by a detailed numerical study.
\par
\begin{figure}[ht]
\vskip 0.3cm
\centerline{\epsfxsize=.77\linewidth\epsfbox{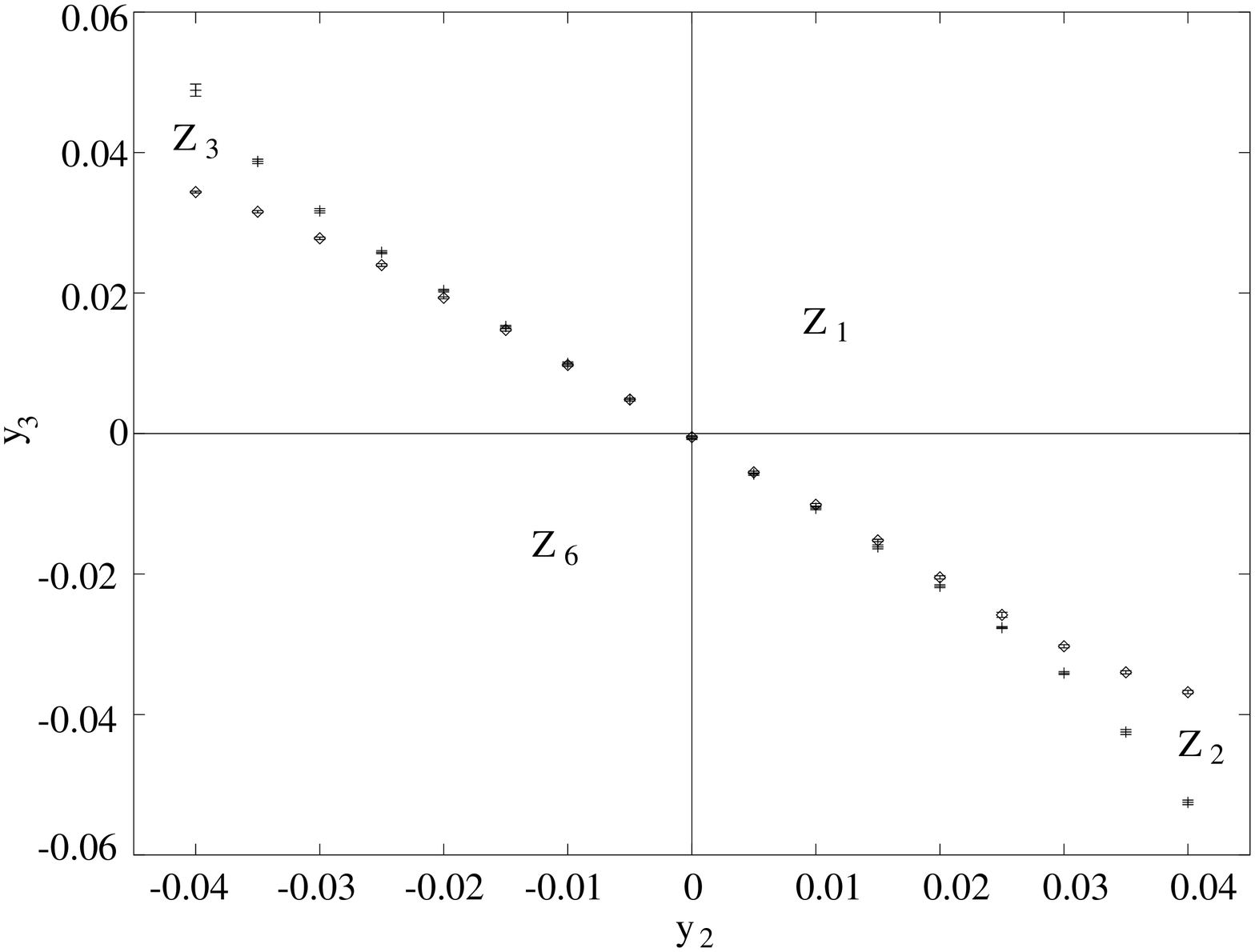}}
\vskip 0.15cm
\centerline{\parbox[t]{0.84\linewidth}{
\small Figure 2: 
Phase diagram on the $K=0.06$ surface.
Phases are labelled according to the nature of their ordering.
Binder cumulants for lattice sizes $L_1=15$ and $L_2=30$ were used.}}
\label{fig2}
\end{figure}
\begin{figure}[ht]
\centerline{\epsfxsize=.77\linewidth\epsfbox{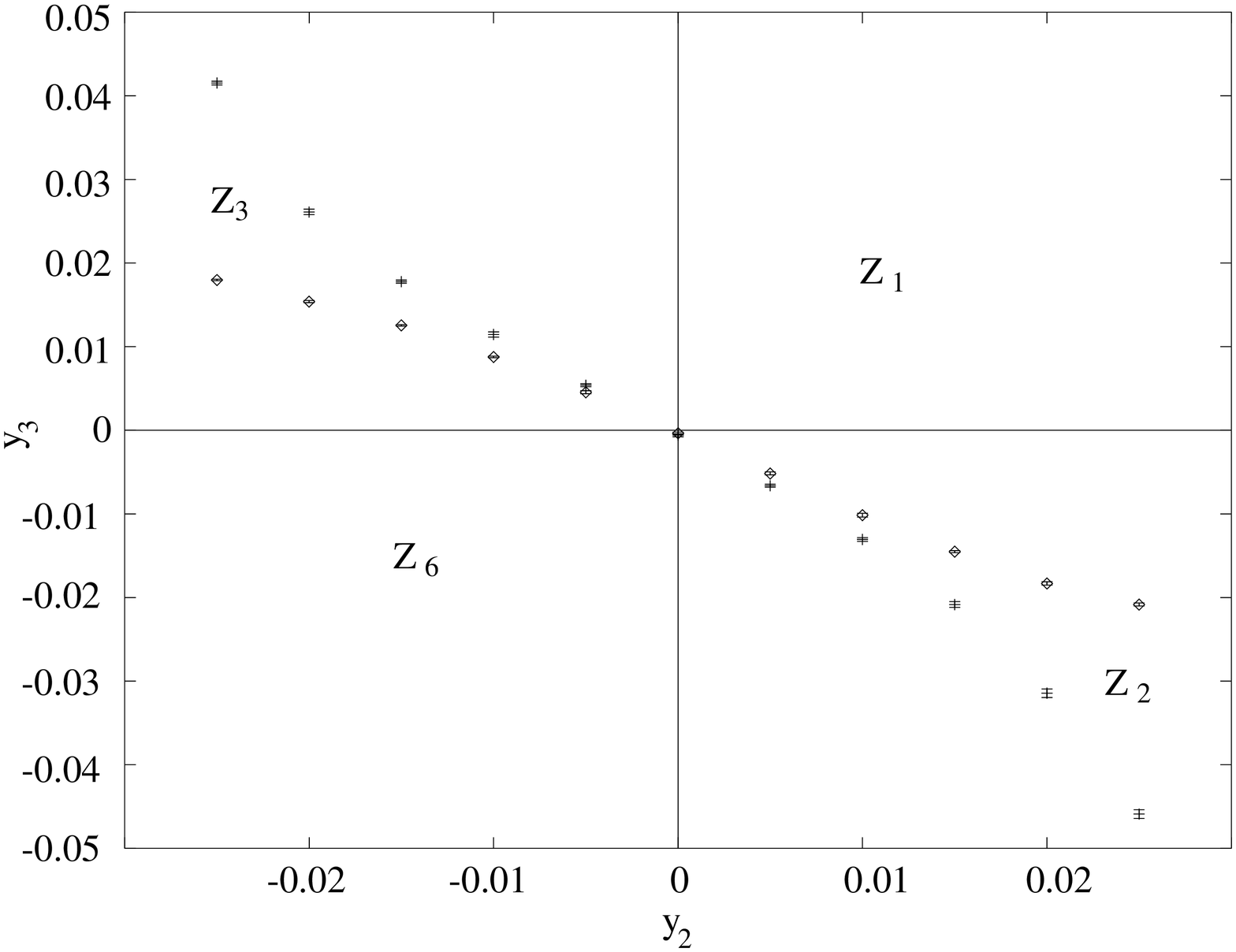}}
\vskip 0.15cm
\centerline{\parbox[t]{0.84\linewidth}{
\small Figure 3: Same as Fig.~2 for $K=0.04$}}
\label{fig3}
\end{figure}
\par
\begin{figure}[ht]
\centerline{\epsfxsize=.77\linewidth\epsfbox{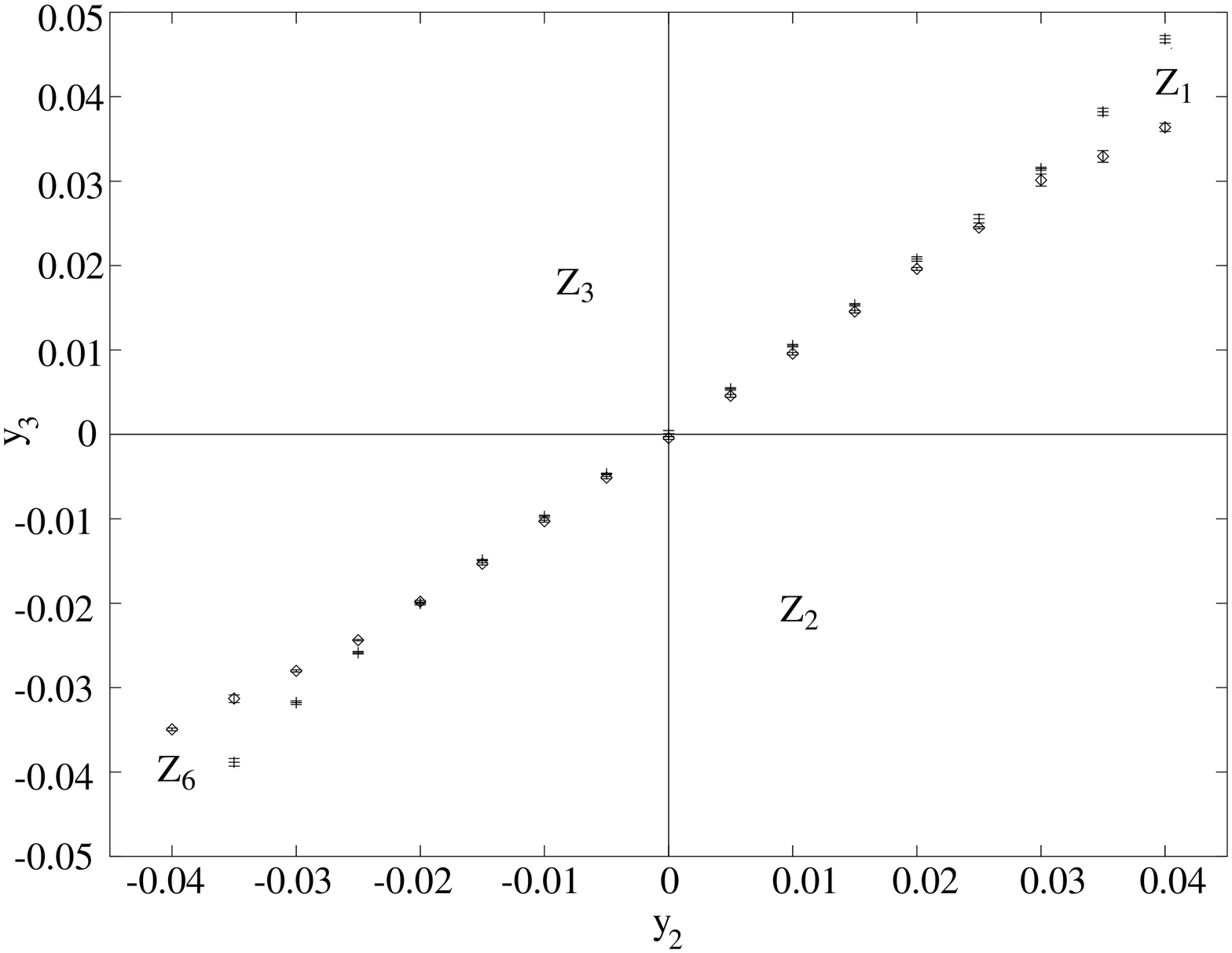}}
\vskip 0.15cm
\centerline{\parbox[t]{0.84\linewidth}{
\small Figure 4: Same as Fig.~2 for $K=-0.08$.
The lattice sizes used here were $L_1=45$ and $L_2=60$.}}
\label{fig4}
\end{figure}
\par

\noindent{\bf Acknowledgements -- }
We would like to thank John Cardy, Michele Caselle, Eytan Domany,
Ferdinando Gliozzi, Martin Hasenbusch, Klaus Pinn and Jean-Bernard Zuber
for useful conversations. The work was supported in part by a TMR 
grant of the European Commission, contract reference ERBFMRXCT960012.
PED thanks the UK EPSRC for an Advanced Fellowship, and RT thanks SPhT
Saclay for hospitality.


\clearpage
\begin{thebibliography}{99}
%
\bibitem{WWa}
F. Y. Wu and Y. K. Wang,
\ttl{Duality transformations in a many-component
 spin model} J. Math. Phys. 17 (1976) 439.
%
\bibitem{DRa}
E. Domany and E. K. Riedel,
\ttl{Two-dimensional anisotropic $N$-vector
 models} Phys. Rev. B19 (1979) 5817.
%
\bibitem{Ca}
J. L. Cardy,
\ttl{General discrete planar models in two dimensions}
 J. Phys. A13 (1980) 1507.
%
\bibitem{AKa}
F. C. Alcaraz and R. K\"oberle,
\ttl{Duality and the phases of $\ZN$ spin
 systems} J. Phys. A13 (1980) L153;
\ttl{The phases of two-dimensional spin and
 four-dimensional gauge systems with $\ZN$ symmetry} J. Phys. A14 (1981)
 1169.
%
\bibitem{GRa}
G. von Gehlen and V. Rittenberg,
\ttl{Hunting for the central charge of the
 Virasoro algebra: six- and eight-state spin models} J. Phys. A19 
 (1986) 2439;
R. Badke,
\ttl{A Monte Carlo renormalisation group study of the
 two-dimensional discrete cubic model: a hint for superconformal
 invariance} J. Phys. A20 (1987) 3425;
G. Sch\"utz,
\ttl{Finite-size scaling spectra in the six-states quantum
 chains} J. Phys. A22 (1989) 731.
%
\bibitem{JKKNa}
J. V. Jos\'e, L. P. Kadanoff, S. Kirkpatrick and D. R. Nelson,
\ttl{Renormalization, vortices, and symmetry-breaking perturbations in
 the two-dimensional planar model} Phys. Rev. B16 (1977) 1217.
%
\bibitem{KLUa}
D. Kim, P. M. Levy and L. F. Uffer,
\ttl{Cubic rare-earth compounds: variants of the three-state Potts model}
Phys. Rev. B12 (1975) 989;
A. Aharony,
\ttl{Critical behaviour of the discrete spin cubic model}
J. Phys. A10 (1977) 389.
%
\bibitem{CPRa}
D. C. Cabra, P. Pujol and C. von Reichenbach,
\ttl{Non-Abelian bosonization and Haldane's conjecture}
Phys. Rev. B58 (1998) 65\hepth{9802014}.
%
\bibitem{EPSa}
S. Elitzur, R. B. Pearson and J. Shigemitsu,
\ttl{Phase structure of discrete Abelian spin and gauge systems}
 Phys. Rev. D19 (1979) 3698.
%
\bibitem{Va}
J. Villain, 
\ttl{Theory of one- and two- dimensional magnets with an easy
 magnetisation plane. II. The planar, classical, two-dimensional
 magnet} J. Physique 36 (1975) 581.
%
\bibitem{FZa}
V. A. Fateev and  A. B. Zamolodchikov,
\ttl{Self-dual solutions of the star-triangle relations in $\ZN$
 models} Phys. Lett. A92 (1982) 37;
A. B. Zamolodchikov and V. A. Fateev,
\ttl{Nonlocal
(parafermion) currents in two-dimensional conformal
 quantum field theory and self-dual critical points in
 $\ZN$-symmetrical statistical systems} JETP 62 (1985) 215.
%
\bibitem{DMSa}
E. Domany, D. Mukamel and A. Schwimmer,
\ttl{Phase diagram of the $\Zf$
 model on a square lattice} J. Phys. A13 (1980) L311;
F. C. Alcaraz,
\ttl{The critical behaviour of self-dual $\ZN$ spin
 systems: finite-size scaling and conformal invariance} J. Phys. A20
 (1987) 2511.
%
\bibitem{DTTa}
P. Dorey, R. Tateo and K. E. Thompson,
\ttl{Massive and massless phases in self-dual $\ZN$ spin models: some
exact results from the thermodynamic Bethe ansatz}
Nucl. Phys. B470 (1996) 317\hepth{9601123}.
%
\bibitem{Ba}
R.J.  Baxter,
 Exactly solved models in Statistical Mechanics,
 Academic Press, London 1982.
%
\bibitem{BWa}
E. Buffenoir and S. Wallon,
\ttl{The correlation length of the Potts model at the first-order
 transition point}
 J. Phys. A26 (1993) 3045.
%
\bibitem{Caa}
J. Cardy, 
\ttl{Effect of random impurities on fluctuation-driven first-order
transitions}
J. Phys. A29 (1996) 1897\condmat{9511112};
A. LeClair, A. W. W. Ludwig and G. Mussardo,
\ttl{Integrability of coupled conformal field theories}
Nucl. Phys. B512 (1998) 523\hepth{9707159}.
%
\bibitem{Za}
A. B. Zamolodchikov,
\ttl{Integrable field theory from conformal field theory}
Adv. Stud. Pure Math. 19 (1989) 641.
%
\bibitem{Pa}
S. Parke,
\ttl{Absence of particle production and factorization of the S matrix in
(1+1)-dimensional models}
Nucl. Phys. B174 (1980) 166.
%
\bibitem{Zb}
C. N. Yang and C. P. Yang,
\ttl{Thermodynamics of one-dimensional system of bosons with repulsive
delta function interaction}
J. Math. Phys. 10 (1969) 1115;
%
Al. B. Zamolodchikov,
\ttl{Thermodynamic Bethe ansatz in relativistic models. Scaling 3-state Potts
and Lee-Yang models}
Nucl. Phys. B342 (1990) 695.
%
\bibitem{RTVa}
F. Ravanini, R. Tateo and A. Valleriani,
\ttl{Dynkin TBAs}
Int. J. Mod. Phys. A8 (1993) 1707\hepth{9207040}.
%
\bibitem{ashkinteller} S. Wiseman and E. Domany, 
\ttl{A cluster method for the Ashkin-Teller model}
Phys. Rev. E48 (1993) 4080\heplat{9310015};
J. Salas and A. D. Sokal, 
\ttl{Dynamic critical behavior of a Swendsen-Wang-type algorithm for 
the Ashkin-Teller model}
J. Stat. Phys. 85 (1996) 297\heplat{9511022}.
%
\bibitem{binder}
K. Binder, 
\ttl{Finite size scaling analysis of Ising model block distribution function}
Z. Phys. B43 (1981) 119;
K. Binder, K. Vollmayr, H.-P. Deutsch, J.D. Reger, M. Scheucher and 
D.P. Landau,
\ttl{Monte Carlo methods for first order phase transitions: some recent 
progress}
Int. J. Mod. Phys. C3 (1992) 1025;
K. Vollmayr, J.D. Reger, M. Scheucher and  K. Binder,
\ttl{Finite size effects at thermally-driven first order phase transitions:
a phenomenological theory of the order parameter distribution} 
Z. Phys. B91 (1993) 113.
%
\end{thebibliography}
\end{document}